\documentclass[preprintnumbers, floatfix, showkeys, preprintnumbers, letterpaper, twocolumn, superscriptaddress,nofootinbib]{revtex4-2}
\pdfoutput=1 
\usepackage{graphicx}
\usepackage{microtype}
\usepackage{amsmath}
\usepackage{amssymb}
\usepackage{subfigure}
\usepackage{url}
\usepackage{hyperref}
\usepackage{mathtools}
\usepackage{orcidlink}

\usepackage{xcolor}
\usepackage{color}
\usepackage{mathrsfs}
\usepackage{calrsfs}
\usepackage{amsfonts}
\usepackage{tabularx}
\usepackage{latexsym}
\usepackage{ragged2e}
\usepackage{epsfig}
\usepackage{textcomp}
\usepackage{float}

\usepackage{caption}
\DeclareCaptionJustification{justified}{\leftskip=0pt \rightskip=0pt \parfillskip=0pt plus 1fil}
\definecolor{vividviolet}{rgb}{0.62, 0.0, 1.0}
\definecolor{amaranth}{rgb}{0.9, 0.17, 0.31}
\definecolor{palatinateblue}{rgb}{0.15, 0.23, 0.89}
\definecolor{brightpink}{rgb}{1.0, 0.0, 0.5}
\definecolor{cornflowerblue}{rgb}{0.39, 0.58, 0.93}
\definecolor{deepcarminepink}{rgb}{0.94, 0.19, 0.22}
\definecolor{radicalred}{rgb}{1.0, 0.21, 0.37}

\hypersetup{ linktoc=all,
	colorlinks, linkcolor={palatinateblue},
	citecolor={brightpink}, urlcolor={amaranth}
}

\graphicspath{{Images/}}

\renewcommand{\d}[1]{\ensuremath{\operatorname{d}\!{#1}}}

\def\sideremark#1{\ifvmode\leavevmode\fi\vadjust{\vbox to0pt{\vss
			\hbox to 0pt{\hskip\hsize\hskip1em
				\vbox{\hsize1.3cm\tiny\raggedright\pretolerance10000
					\noindent #1\hfill}\hss}\vbox to8pt{\vfil}\vss}}}%
\def\beq{\begin{equation}}
\def\eeq{\end{equation}}

\begin{document}
\title{One Gravity, Many $G$'s: Generalized Horizon Entropy and Gravitational Susceptibility}

\author{Yen Chin \surname{Ong}\orcidlink{0000-0002-3944-1693}}
\email{ongyenchin@nuaa.edu.cn}
\affiliation{Center for the Cross-disciplinary Research of Space Science and Quantum-technologies (CROSS-Q), College of Physics, Nanjing University of Aeronautics and Astronautics, \\29 Jiangjun Road, Nanjing City, Jiangsu Province 211106, China}

\begin{abstract}
Effective and thermodynamic derivations of gravity theory in the context of generalized entropy can give rise to multiple, and potentially varying, gravitational couplings, raising the question of which quantity should be identified with the physically measured Newton constant. Motivated by this ambiguity, we formulate the notion of gravitational susceptibility of the horizon.
In spacetimes containing multiple horizons, if the entropy is a function on a multi-horizon state space, $S=S(A_1,A_2,\cdots,A_n)$, then the derivatives of the entropy with respect to individual horizon areas define the horizon gravitational response coefficients. Different horizons may possess distinct gravitational responses without requiring multiple fundamental gravitational couplings, analogous to direction-dependent response coefficients in macroscopic media. This framework naturally suggests the possibility that horizon entropies (such as the cosmological apparent horizon and a black hole horizon) can influence one another, which we explore as a possible phenomenological route to the black hole-cosmology coupled ``black hole mass growth'' previously discussed in the literature, without claiming to derive its observed scaling. 
\end{abstract}

\maketitle

{
\begin{flushright}
\begin{minipage}{0.8\columnwidth}
\itshape
``We confine ourselves to weak perturbation and ask for the
response of the system in the linear approximation.''

\hfill --- Ryogo Kubo
\end{minipage}
\end{flushright}
}

\section{Introduction: Horizon-Dependent Gravitational ``Constant''}

Jacobson's seminal work \cite{9504004} essentially shows that the form of entropy being a linear function of the horizon area gives rise to general relativity. This in turn implies that if one were to generalize horizon entropy beyond the standard area law, from $S=A/4G$ to $S=f(A)/4G$, then the gravitational theory would necessarily need to be revised. In other words, one should not keep the gravity theory as general relativity (GR), and simply replace the Bekenstein-Hawking area law by another entropy expression on top of that. Following Jacobson's approach, it was shown in \cite{2407.00484} that the entropy modification above would keep the \emph{form} of Einstein's field equation intact, but Newton's constant is modified by
\begin{equation}\label{Geff}
G \mapsto G_\text{eff}=\frac{G}{f'(A)}. 
\end{equation}
That is to say, $G$ is not strictly a constant if the horizon area changes. (See also the recent discussion in \cite{2602.20430}, and some implications explored in \cite{2505.03907,2505.07972}.)
The derivation follows Jacobson's method closely. In particular, due to the matter flux
\begin{equation}
\delta Q = - \kappa \int_\mathcal{H} \lambda T_{ab} k^a k^b \text{d}\lambda \text{d}A
\end{equation}
through a horizon,
the change of the horizon area is given by
\begin{equation}
\delta A = - \int_\mathcal{H} \lambda R_{ab} k^a k^b \text{d}\lambda \text{d}A,
\end{equation}
where $\lambda$ is an affine parameter and $k^a$ is the tangent null vector to the horizon generators. 

The entropy variation is $\delta S=\frac{f'(A)}{4G} \delta A$. Using the Clausius relation $\delta Q=T\d S$, and identifying the temperature with essentially the surface gravity, $T=\kappa/2 \pi$, we obtain
\begin{equation}
-\kappa \int \lambda T_{ab} k^a k^b \text{d}\lambda \text{d}A = - \frac{\kappa}{2\pi} \int \lambda \frac{f'(A)}{4G}  R_{ab} k^a k^b \text{d}\lambda \text{d}A.
\end{equation}
This relation must hold for all null vectors $k^a$; therefore, the corresponding integrands must agree. This implies
\begin{equation}
T_{ab} k^a k^b = \frac{f'(A)}{8\pi G} R_{ab} k^a k^b.
\end{equation}
Finally one obtains the Einstein-like field equations
\begin{equation}
R_{ab} - \frac{1}{2}R g_{ab} + \Lambda g_{ab} = 8\pi G_{\text{eff}} T_{ab},
\end{equation}
with $G$ replaced by Eq.(\ref{Geff}). This was referred to as the ``generalized entropy and varying-G'' (GEVAG) framework \cite{2407.00484}.

Two major problems naturally arise: (1) what is the ``$G$'' one measures in laboratory experiments? (2) What if there are multiple horizons? For example, given two or more black holes, they each have a different area, and thus different $G_\text{eff}$. What does that mean? Some attempts were already made to interpret these puzzles in the original GEVAG paper \cite{2407.00484}, but the discussions therein were rudimentary. 

It should be emphasized that usually Jacobson's method is understood as a \emph{local} derivation, since only a local patch of a Rindler horizon is required for it to work. However, this is due to the specialness of GR. The derivation itself requires an area integral, but the area dependence ultimately drops out. Therefore, in GR, the field equation thus obtained is not dependent on the horizon area, and therefore can be applied in spacetimes with any number of horizons. In GEVAG, however, the area dependence is inherited by the final field equation. This in turn suggests that in order to answer what happens in the multi-horizon case, we need to consider \emph{ab initio} a multi-horizon setup.

It should be emphasized that the presence of multiple null surfaces in a spacetime does not necessarily imply a multi-horizon thermodynamic system. For example, in Kerr and Reissner-Nordstr\"om black holes, although an inner and an outer horizon exist (putting aside the instability issue of the inner horizon), the standard black hole thermodynamics is associated with the outer event horizon only. The inner horizon is not treated as an independent thermodynamic boundary, and the entropy modification can therefore be consistently formulated using the original single-horizon GEVAG approach. The multi-horizon extension considered here instead refers to situations in which the causal region of interest is bounded by multiple physically independent null components, such as a black hole horizon together with a cosmological horizon, or multiple black hole horizons.

The structure of the present paper is as follows: in Sec.(\ref{II}), we extend Jacobson's method to GEVAG in the multi-horizon case and argue that this leads to multiple $G$'s, each for one horizon. This is expected if the effective gravitational coupling is area dependent. Then in Sec.(\ref{III}) we argue that this is not really a problem; rather we should re-interpret what ``multiple $G$'s'' means. We do so by proposing the idea of ``gravitational susceptibility'' and point out why this point of view naturally suggests the possibility that various horizons may couple, leading in particular to black hole-cosmology coupling explored in Sec.(\ref{IV}). After that in Sec.(\ref{V}), we discuss the effect of including a dissipative term, which is natural when the thermodynamical system is not in thermal equilibrium. We end in Sec.(\ref{VI}) with more discussions on the implications of this new framework, and propose some future research directions that can be pursued. 

\section{Multi-Horizon Spacetimes Imply Multiple Effective G?}\label{II}

For simplicity, let us begin with the two-horizon case. Consider a causal patch region $\mathcal{P}$ whose boundary contains two null components $\partial \mathcal{P} = \mathcal{H}_1 \cup \mathcal{H}_2$. We associate the cross-sectional areas $A_1$ and $A_2$ with a joint entropy $S_{\mathcal{P}}=S(A_1,A_2)$. Here the entropy describes the joint causal patch; the special case of independent horizons is recovered when $S(A_1,A_2)=S_1(A_1)+S_2(A_2)$. Nonseparable contributions are an additional physical possibility, for instance from shared constraints or correlations, not an inevitable consequence of having two horizons. 
We should expect in general that if the system is not in equilibrium, the entropy variation should take the form
\begin{flalign}
\delta S_\mathcal{P} &= S'_1 \delta A_1 + S'_2 \delta A_2\\
&= \frac{\delta Q_1}{T_1} + \frac{\delta Q_2}{T_2} + \delta_\text{int} S.
\end{flalign}
where $S'_i \equiv \partial S/\partial A_i$,  
and $\delta_\text{int} S$ represents an internal entropy production term \cite{0602001,0909.4194}. For now, we will make the assumption that this term is zero (so that we are in a reversible equilibrium scenario). We will also assume that the areas can be independently varied. These assumptions provide a simple starting point for our analysis. This is an assumption about admissible virtual variations, not about whether $S$ is separable. When the variations are constrained, the individual terms in the summed Clausius relation need not vanish separately, and the horizon-by-horizon identification below is then conditional.

Depending on whether matter fluxes enter or leave the patch through each of the $A_i$'s, one could write 
\begin{equation}
\delta Q_i = -\epsilon_i \int_{\mathcal{H}_i} \kappa_i \lambda_i T_{ab} k_i^a k_i^b \d\lambda_i \d A_i,
\end{equation}
where $\epsilon_i = 1$ for energy entering the patch, and $\epsilon_i = -1$ for energy leaving the patch.
The associated change in the individual area is then
\begin{equation}
\delta A_i = -\epsilon_i \int_{\mathcal{H}_i}  \lambda_i R_{ab} k_i^a k_i^b \d\lambda_i \d A_i.
\end{equation}
We therefore have the relation
\begin{equation}
\sum_{i=1}^2 \epsilon_i \int_{\mathcal{H}_i} \lambda_i \left(S'_i R_{ab} - \frac{2\pi}{\hbar}T_{ab}\right)k_i^ak_i^b \d\lambda_i \d A_i = 0,
\end{equation}
where we have used the temperature defined by surface gravity: $T_i = \hbar \kappa_i/2\pi$. We otherwise use units $\hbar=1$, as in Eqs.~(4) and (5). 

As mentioned before, we first assume that the individual areas can be varied independently. For example, one may consider injecting a localized pulse of radiation into a black hole. This should not affect the size of another black hole far away. It is interesting to ask what happens if this assumption is relaxed. The entropy coupling introduced below does not itself imply such dependent area variations. This in turn can provide a theoretical framework to explain why there might be a ``cosmology-black hole coupling'', in which the mass of the black holes may change as the Universe expands, as previously discussed in the literature \cite{2212.06854,2302.07878,2306.08199,2307.10708,2307.02474,2312.12344,2405.12282,2306.11588,2407.14549,2410.10459,2504.20338,2601.03296,2607.13857,2603.24609} (see, however, \cite{2507.03408}). For now, we consider the simpler case in which the areas can be varied independently. This would imply that 
\begin{equation}
S'_i R_{ab} k_i^a k_i^b = \frac{2\pi}{\hbar} T_{ab}k_i^a k_i^b
\end{equation}
holds for each $i$.
Consequently, the GEVAG construction applies to each horizon separately. Therefore, to each horizon $A_i$, we associate a distinct effective gravitational ``constant'':
\begin{equation}
G_\text{eff}^i = \frac{1}{4S'_i}.
\end{equation}
If we write $S=f(A)/4G$ as in the original GEVAG paper, then $G_\text{eff}^i = {G}/{f_i'(A_i)}.
$

In other words, for a multi-horizon spacetime with entropy $S=S(A_1,A_2,\cdots A_n)$, the Einstein-like field equation is schematically of the form
\begin{equation}
R_{ab} - \frac{1}{2}g_{ab}R + \Lambda g_{ab}= 8\pi \mathcal{G}T_{ab},
\end{equation}
where 
\begin{equation}
\nabla_A S =\left(\frac{\partial S}{\partial A_1}, \frac{\partial S}{\partial A_2}, \cdots, \frac{\partial S}{\partial A_n}\right)
\end{equation}
is the \emph{entropy gradient} vector, and $\mathcal{G}$ is a map from the vector to a horizon-associated response coefficient for each $A_i$, rather than a uniquely determined local bulk coupling. That is to say, 
\begin{equation}
\mathcal{G}: \nabla_A S \longmapsto \frac{1}{4}\left(\frac{\partial S}{\partial A_i}\right)^{-1}.
\end{equation}
Here $\Lambda$ denotes the same integration-constant term as in Eq.~(6). In single-horizon GEVAG, $G_{\rm eff}(A)$ is area-dependent, not an independently dynamical scalar field. With an Einstein-form field equation and constant $\Lambda$, the Bianchi identity requires $\nabla^a(G_{\rm eff}T_{ab})=0$, so matter need not be separately conserved \cite{2603.23551}. In the multi-horizon case, the equation remains schematic: it does not specify how distinct horizon responses enter a unique bulk dynamics, a question to which we return in Sec.\eqref{VI}.

This result seems disastrous at first since it seems to imply that all nonlinear extensions of the Bekenstein-Hawking area law would yield distinct gravitational constants for each black hole. However, this may not be as strange as it first appears to be. Recall that gravitational constant itself is not an observable; it is always tied to masses. That is, as already pointed out in \cite{2407.00484}, changing a constant $G$ to a horizon-dependent $G_\text{eff}$ will appear to an observer as a change in the mass of the black hole. In the multi-horizon case, the same interpretation holds: there are many $G_\text{eff}$'s, one for each black hole, but we should not interpret them as fundamental gravitational couplings. They are just emergent quantities. In fact, an analogy to dielectric media can shed some light on the situation at hand.

\section{Gravitational Response and Gravitational Susceptibility}\label{III}

As an analog, consider the dielectric displacement in electromagnetism:
\begin{equation}
D = \epsilon_0 E + P,
\end{equation}
where $E$ denotes the electric field and $P$ the polarization of the material. The fundamental constant here is $\epsilon_0$, the vacuum permittivity that enters Maxwell's equations. However, if the polarization is that of a linear dielectric, then
\begin{equation}
P = \chi_e \epsilon_0 E,
\end{equation}
for some susceptibility $\chi_e$. Consequently,
\begin{equation}
D = \epsilon E,
\end{equation}
where $\epsilon=\epsilon_0 (1+\chi_e)$ is the dielectric ``constant''. It is not a new fundamental coupling, but a macroscopic response coefficient that describes how the material reacts in the presence of an electric field $E$. Essentially, $\chi_e \propto \partial P/\partial E$. In this language, we can view the gravitational constant in a new light, namely it is related to the \emph{gravitational susceptibility}
\begin{equation}
\chi_{g} \equiv \frac{\partial S}{\partial A}.
\end{equation}
One may also call this quantity the \emph{gravitational response coefficient}, just like the $\chi_e$ above measures electric polarization response.

{Kubo's observation in the opening epigraph \cite{kubo} motivates the analogy with linear response; here the response is an entropy-area derivative, not a microscopic Kubo correlation-function formula.}\par

In the linear response approximation, we have 
\begin{equation}
\delta S \approx \chi_g(A_0)\delta A,
\end{equation}
for some initial area $A_0$ and
\begin{equation}
S(A_0+\delta A)=S(A_0) + \left.\left(\frac{\partial S}{\partial A}\right)\right\rvert_{A_0}\delta A + O(\delta A^2).
\end{equation}
The linear approximation here does not mean that the entropy function itself is linear. It also does not mean that we are in the weak-field or weak gravity regime (the small quantity is $\delta A$, not the metric perturbation).

In GR, $\chi_{g}$ is just $1/4G$, but in GEVAG, we would have 
\begin{equation}
\chi_{g} = \frac{1}{4G_\text{eff}}.
\end{equation}
This follows from a direct calculation: since the generalized entropy is $S=f(A)/4G$, its variation is
\begin{equation}
\delta S = \frac{f'(A)}{4G} \delta A = \frac{1}{4G_\text{eff}}\delta A,
\end{equation}
so that in GEVAG,
\begin{equation}
 \frac{\partial S}{\partial A} = \frac{1}{4G_\text{eff}}.
\end{equation}
Note that this is \emph{not} equivalent to saying that the entropy is the form $A/4G_\text{eff}$. This is not equivalent as was discussed in Sec.(\ref{IV}) of \cite{2407.00484}.

Just like a linear dielectric displacement $D=\epsilon E$, GR has a linear entropy function $S=A/4G$, giving rise to a constant response. Generalized entropy with $G_\text{eff}=G_\text{eff}(A)$ is thus similar to a medium for which $\epsilon=\epsilon(E)$. The horizon ``medium'' changes its response as its area changes. It is therefore natural to consider the entropy gradient, since the response is no longer described by a single global number. This is similar to some media requiring a dielectric tensor to describe its response in different directions: $D_i = \sum_{j}\epsilon_{ij}E_j$.

Now let us allow the entropy to be nonseparable while retaining independently variable areas. Previously, with this assumption in place, we obtain $\chi_g^i = S'_i \equiv \partial S/\partial A_i$ essentially from the first derivative information contained in $\partial S$. To encode how one horizon responds when another horizon is perturbed, we would need the second derivatives. The natural thing to do is therefore to consider the equation
\begin{equation}
\delta S'_i = \sum_j \frac{\partial S'_i}{\partial A_j} \delta A_j.
\end{equation}
This is a matrix equation. The central object of interest is $\frac{\partial S'_i}{\partial A_j}$, or more explicitly written,
\begin{equation}
H_{ij} \equiv \frac{\partial^2 S}{\partial A_i \partial A_j}.
\end{equation}
This Hessian of the entropy is a ``horizon coupling matrix'' that would tell us whether the two horizons are interdependent, in the sense that the gravitational susceptibility of one is a function of another. For example, if the entropy is linear and additive, like in GR, then $S= S_1 + S_2 + \cdots + S_n$, and $H_{ij}\equiv 0$ for all values of $i$ and $j$. However, if, just for an example, a two-horizon spacetime has entropy of the form
\begin{equation}
S = \frac{A_1+A_2}{4G} + \mu A_1^\alpha A_2^\beta,
\end{equation}
then the horizon coupling matrix is nonzero. For smooth $S$, $H_{ij}=H_{ji}$; equivalently, $\partial_j(1/G_{\mathrm{eff}}^i)=\partial_i(1/G_{\mathrm{eff}}^j)$. This reciprocity condition constrains horizon responses arising from a common entropy function. The susceptibility language organizes these entropy-derivative relations rather than introducing independent microscopic physics.

In addition, let us recall that the heat flow equation (Clausius relation) $\delta Q = TdS$ enters Jacobson's argument. Previously we assumed independently variable areas, which results in 
\begin{equation}
\frac{\delta Q_1}{T_1} + \frac{\delta Q_2}{T_2} = \delta S_1 + \delta S_2 = S'_1 \delta A_1 + S'_2 \delta A_2.
\end{equation}
In other words, 
\begin{equation}
\frac{\delta Q_1}{T_1} + \frac{\delta Q_2}{T_2} = \chi_1 \delta A_1 + \chi_2 \delta A_2.
\end{equation}
Here $\delta S_i$ denotes the contribution $S'_i\delta A_i$, not necessarily the variation of a separate entropy $S_i(A_i)$. For coupled horizons, $\chi_1$ is no longer a function of just $A_1$ but also that of $A_2$.
However, in this case there are off-diagonal terms $H_{12}$ and $H_{21}$ in the entropy Hessian that describe how a perturbation of one horizon induces a response in the other horizon. The horizons are coupled in, and only in, this sense. In particular, $H_{12}\neq0$ does not require $\delta A_1$ and $\delta A_2$ to be dependent.

For example, if hypothetically there exists an entropy of the form
\begin{equation}
S=\frac{1}{4G}(A_1+A_2 + \mu A_1 A_2),
\end{equation}
then 
\begin{equation}
\chi_{1} = \frac{\partial S}{\partial A_1} = \frac{1}{4G}(1+\mu A_2),
\end{equation}
and thus
\begin{equation}
G_\text{eff}^1 = \frac{G}{1+\mu A_2},
\end{equation}
so the coupling changes the black hole response. In other words, $G_\text{eff}^1$ is now ``environment-dependent'':\footnote{{This is reminiscent, but only vaguely so, of the old ``Mach principle''.}} the gravitational response of the first horizon depends on the area of the second horizon. 
The coupling is off-diagonal since $H_{11} = H_{22}=0$ but
\begin{equation}
H_{12}=H_{21}= \frac{\partial^2 S}{\partial A_1\partial A_2} = \frac{\mu}{4G}\neq 0.\end{equation}

We should be careful of the interpretation here, however. Nonzero $H_{ij}$ does not indicate that there is a statistical correlation between the various horizons. It is also not a ``dynamical coupling'' in the stronger sense that one has coupled time evolution of the horizons, i.e., $\dot{A}_i=F(A_1,A_2,\cdots, A_n)$. All we can say from the thermodynamic viewpoint is that the gravitational susceptibility of one horizon is a function of the areas of other horizons. In other words, the response itself does not dictate the full dynamics and the subsequent geometrical evolutions (see \cite{discourse} for a more detailed discussion on how much geometry does generalized entropy determine). 

\section{Black Hole-Cosmology Coupling}\label{IV}

Consider a possible coupling between the cosmological apparent horizon entropy $S_c$, and a black hole entropy $S_b$, in the form\footnote{The minus sign in front of $\gamma$ is just a convention, so that for $\gamma>0$, $G_b$ increases as the cosmological horizon increases. One should of course choose $\gamma$ such that $G_b$ does not become negative.}
\begin{equation}\label{model}
S(A_b, A_c) = \frac{1}{4G}\left(A_b + A_c - \gamma \frac{A_bA_c}{A_0}\right),
\end{equation}
where $A_0$ is a reference area (so that $\gamma$ is dimensionless). For example, it could be the initial area of the cosmological horizon at some chosen time. The cross term is a lowest-order phenomenological ansatz, not derived from horizon microphysics. $A_0$ is held fixed; only $\gamma/A_0$ enters the model, so choosing $A_0=A_c(a_i)$ is a convenient convention. Its applicability must be checked in any concrete cosmology.

The gravitational response is
\begin{equation}
\frac{\partial S}{\partial A_b} = \frac{1}{4G}\left(1-\gamma\frac{A_c}{A_0}\right),
\end{equation}
which according to our preceding discussion, means that the effective gravitational ``constant'' for the black hole is a function of the cosmological horizon area. Explicitly, it is given by 
\begin{equation}\label{Gb_response}
G_b(A_c) = \frac{G}{1-\gamma\frac{A_c}{A_0}}.
\end{equation} 
Likewise, by symmetry, we have
\begin{equation}
G_c(A_b) = \frac{G}{1-\gamma\frac{A_b}{A_0}}.
\end{equation}
The Hessian satisfies
\begin{equation}
H_{bb}=\frac{\partial^2 S}{\partial A_b^2} = 0; ~~H_{cc}=\frac{\partial^2 S}{\partial A_c^2} = 0,
\end{equation}
but
\begin{equation}
H_{bc} = \frac{\partial^2S}{\partial A_b \partial A_c} = -\frac{\gamma}{4GA_0},
\end{equation}
signifying there is a coupling between the two horizons. For $G>0$, finite positive responses require $1-\gamma A_c/A_0>0$ and $1-\gamma A_b/A_0>0$. For $\gamma>0$ these exclude the poles at unity; a cosmological trajectory approaching such a pole leaves the physical domain of this ansatz, whose validity cannot be assumed near that regime. For $\gamma\leq0$ with positive areas these particular poles do not occur.

In the phenomenological interpretation adopted here, suppose the black-hole geometry is characterized by the gravitational parameter $G_b\mathcal{M}$, with $\mathcal{M}$ a constant mass parameter. An observer who assumes the reference coupling $G$ would instead infer a mass $M_\text{eff}$ from $G M_\text{eff}=G_b\mathcal{M}$. Using Eq.~(\ref{Gb_response}), this gives
\begin{equation}
M_\text{eff}(a) = \frac{\mathcal{M}}{1-\gamma\frac{A_c}{A_0}},
\end{equation} 
where $M_\text{eff}$ is an inferred gravitational mass, not by itself evidence of physical accretion or growth of a conserved mass.

Let us define the ``effective mass-growth index'' 
\begin{equation}
k(a) \equiv \frac{\mathrm{d}\ln M_\text{eff}}{\mathrm{d}\ln a}.
\end{equation}
This implies, upon integration,
\begin{equation}
M_\text{eff}(a) = M_i \exp \left(\int_{a_i}^a k(\tilde{a}) \frac{\text{d}\tilde{a}}{\tilde{a}} \right),
\end{equation}
where $M_i$ is the initial mass. Henceforth, the subscript $i$ denotes the initial value of the corresponding quantity.

If $k$ is a constant, the integral gives 
\begin{equation}
\ln \left(\frac{M_\text{eff}}{M_i}\right) = k \ln \left(\frac{a}{a_i}\right),
\end{equation}
or, explicitly, the effective mass as a function of the scale factor is
\begin{equation}\label{Meff}
M_\text{eff}(a) = M_i \left(\frac{a}{a_i}\right)^k.
\end{equation}
This result approximately holds if $k$ is slowly varying instead of being exactly constant. This is the form of black hole-cosmology coupling investigated in the cosmology and astrophysics literature. However, note that this \emph{form} of the equation follows directly from the effective mass-growth index \emph{definition} and has nothing to do with entropy yet. The entropy is only needed when we need to compute explicitly what $k$ is.
Writing $x(a):=\gamma A_c(a)/A_0$, the entropy ansatz gives explicitly
\begin{equation}
k(a)=\frac{x(a)}{1-x(a)}\frac{\mathrm{d}\ln A_c}{\mathrm{d}\ln a}.
\end{equation}
For a spatially flat FLRW apparent horizon $A_c=4\pi/H^2$ (in units $c=1$), this is $k=-2x(1-x)^{-1}\mathrm{d}\ln H/\mathrm{d}\ln a$. For example, on a prescribed dust-dominated background with $A_c\propto a^3$, one obtains $k=3x/(1-x)$. The value $k=3$ occurs at $x=1/2$, but $\mathrm{d}k/\mathrm{d}\ln a=18$ there: the value is not approximately constant in that example. This background is illustrative, not a self-consistent solution of modified field equations.

As a second diagnostic, consider a prescribed spatially flat $\Lambda$CDM expansion history $H^2=H_0^2(\Omega_{m0}a^{-3}+\Omega_{\Lambda0})$. Define $u(a)=(\Omega_{m0}/\Omega_{\Lambda0})a^{-3}$ and $\xi=\gamma A_\infty/A_0$, with $A_\infty=4\pi/(H_0^2\Omega_{\Lambda0})$ the asymptotic apparent-horizon area. Then $A_c=A_\infty/(1+u)$ and
\begin{equation}
k(a)=\frac{3\xi u(a)}{[1+u(a)][1+u(a)-\xi]}.
\end{equation}
For $0<\xi<1$, $G_b$ stays finite and positive along this prescribed history, but $k\to0$ at late times rather than approaching a sustained value of $3$. To determine the maximum, we differentiate $k$ with respect to $u$:
\[
\frac{\mathrm{d}k}{\mathrm{d}u}=\frac{3\xi(1-\xi-u^2)}{(1+u)^2(1+u-\xi)^2}=0,
\]
which gives $u=\sqrt{1-\xi}$ on the physical branch $u>0$. Substituting this stationary point into the preceding expression for $k$ yields
\begin{equation}\label{kmax}
k_{\rm max} = 3\frac{1-\sqrt{1-\xi}}{1+\sqrt{1-\xi}}<3.
\end{equation}
Furthermore, $A_c=A_\infty/(1+u)$ and Eq.~(\ref{Gb_response}) give $G_b/G=[1-\xi/(1+u)]^{-1}$. At the same stationary point, this becomes
\begin{equation}\label{Gb_at_kmax}
\left.\frac{G_b}{G}\right|_{k_{\rm max}}=\frac{1}{\sqrt{1-\xi}}.
\end{equation}
Thus $k$ can approach $3$ on the regular branch only at the price of an increasingly large effective response. The limiting choice $\xi=1$ has a different asymptotic behavior, to which we return in Sec.~\ref{VI}. Neither background is derived as a solution of the coupled GEVAG equations.

We can see how, in principle, this can give rise to an effective dark energy. Let us consider a population of black holes with a conserved comoving number. That means the number density of black holes is
\begin{equation}\label{blackholedensity}
n_b (a) = n_{b,i} \left(\frac{a}{a_i}\right)^{-3}.
\end{equation}
The effective black hole energy density is
\begin{equation}
\rho_b(a) = n_b(a)M_\text{eff}(a).
\end{equation}
With Eq.(\ref{Meff}) for $M_\text{eff}$, we obtain
\begin{equation}
\frac{\mathrm{d} \ln \rho_b}{\mathrm{d} \ln a} = -3 + k(a).
\end{equation}

In standard FLRW cosmology in GR, the continuity equation can be written as
\begin{equation}
\frac{\mathrm{d} \ln \rho}{\mathrm{d}\ln a} = -3(1+w),
\end{equation}
where $w$ is the equation of state parameter $w=p/\rho$.
This suggests, in our case, the effective interpretation
\begin{equation}
w_\text{eff}(a) = -\frac{k(a)}{3}.
\end{equation}
This is a scaling-based effective equation-of-state parameter, not one obtained by imposing a separately conserved physical stress-energy tensor.

In particular, $k=3$ yields $w_\text{eff}(a) =-1$. This recovers the claim in the black hole-cosmology coupling literature that $k=3$ in Eq.(\ref{Meff}) mimics a cosmological constant. However, the present entropy ansatz neither predicts $k=3$ nor ensures that $k$ remains approximately constant; an appropriate evolution of $A_c(a)$ must be supplied. In principle, given a fixed $k$, we can solve for $\gamma$ in terms of $A_c$ and its time-evolution. 

However, we need to emphasize that whether the expansion of the Universe accelerates needs to be analyzed carefully in our setting due to the varying gravitational coupling. In particular, the Friedmann equation would be different \cite{2603.23551}. In addition, even for normal matter like dust, the continuity equation is not zero on the RHS, but rather $-\rho \dot{G}_\text{eff}/G_\text{eff}$ \cite{2407.00484,2603.23551} (Eq.(\ref{blackholedensity}) above for \emph{number} density can hold even though energy density does not). As a consequence, we \emph{cannot} simply write, as in GR, $\rho \propto a^{-3(1+w_\text{eff})}$. Thus, even though for $1 < k < 3$ the effective black-hole density has the same scale-factor dependence as a conventional quintessence fluid in GR, its actual dynamics would require a more careful analysis. Since this work is only meant as a demonstration of the theoretical framework, we leave the full cosmological analysis for a separate future work.

\section{The Effect of the Internal Entropy Production Term}\label{V}

As mentioned in Sec.(\ref{II}), we expect that in general, when the gravitational system is out of equilibrium, there should be an entropy production term $\delta_\text{int} S$ such that
\begin{equation}
\delta S_\mathcal{P} = \frac{\delta Q_1}{T_1} + \frac{\delta Q_2}{T_2} + \delta_\text{int} S.
\end{equation}
Thus far, we have not considered this term. What is its effect?

We first examine whether the gravitational susceptibility relation remains valid, namely, whether
\begin{equation}
\frac{\partial S}{\partial A_i} = \frac{1}{4G_\text{eff}^i}.
\end{equation}
This identification remains unchanged at first order if the internal entropy production does not contribute at first order in the admissible $\delta A_i$, as can happen in near-equilibrium treatments where entropy production begins at higher perturbative order. Dependence on the background areas alone does not establish this condition.

If, however, there is a dependence on $\delta A_i$, e.g., 
\begin{equation}
\delta_\text{int} S= \sum_i \Pi_i \delta A_i,
\end{equation}
the Clausius relation becomes
\begin{equation}
\frac{\partial S}{\partial A_i} \delta A_i = \frac{\delta Q_i}{T_i} + \Pi_i \delta A_i,
\end{equation}
or equivalently, with our notation that $S'_i \equiv \partial S/\partial A_i$,
\begin{equation}
\frac{\delta Q_i}{T_i} = \left(S'_i-\Pi_i\right)\delta A_i.
\end{equation}
Consequently, the effective gravitational constant is
\begin{equation}\label{ent2}
G_\text{eff}^i = \frac{1}{4(S'_i-\Pi_i)}.
\end{equation}
Along an actual evolution, the second law requires $\dot S_{\rm int}=\sum_i\Pi_i\dot A_i\geq0$; it does not require every $\Pi_i$ to be positive. Since the ansatz is linear in the area variations, it cannot give nonnegative entropy production for arbitrary independently reversible perturbations without additional constitutive restrictions on the admissible evolution.
If we interpret $S'_i$ as the equilibrium gravitational susceptibility of a horizon with area $A_i$, then $\Pi_i$ can be viewed as a dissipative correction associated with entropy production. A finite positive response requires $S'_i-\Pi_i>0$. The pole at equality signals a breakdown of this effective response formula, not an established spacetime singularity.
The quantity $G_\text{eff}^i$ is then the gravitational response after relaxation of the corresponding ``medium''; cf. Debye relaxation for dielectric media.

In general, therefore, the relation
\begin{equation}
G_\text{eff}^i \propto \frac{1}{S'_i}
\end{equation}
is only the equilibrium limit of a more general dissipative response theory. Interestingly Eq.(\ref{ent2}) above suggests that for $0<\Pi_i<S'_i$ (for the same $S'_i$), a larger value for $G_\text{eff}^i$ is possible. In other words, a horizon out of thermodynamic equilibrium can gravitate more strongly than the equilibrium horizon within this phenomenological description; the second law must still be checked along its evolution. 

We discuss further possible implications of the irreversible gravitational entropy production term in the next section. For now, we should note that the description in this section is on a phenomenological ``response-level'', given that we have interpreted $G_\text{eff}$ as a gravitational response. However, if we were to derive the susceptibility framework starting again from Jacobson's method, we should expect $f(R)$-type gravity since it is already known that an internal entropy production term gives rise to $f(R)$ gravity (before we consider generalized entropy) \cite{0602001}. This does not necessarily mean that the two approaches give different results (though of course one has to construct the full multi-horizon $f(R)$-type theory first to compare; this is beyond the scope of the current work). Rather, $\Pi_i$ only gives us the gravitational response ``re-scaling'', whereas the full construction could reveal more the underlying dynamics, i.e., $\Pi_i$ could be derived from spacetime variation of some generalized entropy density.
(See also \cite{2602.20430} for the relation between $f(R)$ gravity and curvature-dependent entropy density.)  

\section{Discussion: Implications of the New Framework}\label{VI}

In this work, we have proposed that the entropy-area susceptibility $\chi_i = \partial S/\partial A_i$ plays the role analogous to a constitutive response\footnote{This concept of ``response'' is different from the one discussed in \cite{2608.07046} in the context of the entropy-gravity correspondence. In that work, the ``response'' refers to a response to a source. Roughly speaking the generalized entropy determines a nonlocal kernel, and gravity responds to this dressed source.} coefficient in material systems. Then the various 
\begin{equation}
G_\text{eff}^i \sim \left(4\frac{\partial S}{\partial A_i}\right)^{-1}
\end{equation}
are different gravitational response coefficients of the same spacetime, not several fundamental Newton constants. We need not be concerned about different $G_\text{eff}$ tied to different horizons, any more than we need to worry about the existence of several dielectric responses in an anisotropic dielectric medium (and neither of them is the ``actual'' permittivity). GR is special as in that case the response is universal, so all the response coefficients are the same fundamental constant. The most natural possibility, as proposed in this work (as well as in the original GEVAG paper \cite{2407.00484}), is that $G_\text{measured}$ is associated with the cosmological horizon. This remains a hypothesis: a local weak-field calculation matching $G_c$ to a laboratory measurement has not been carried out, and a combination of responses is not excluded. In other words, what we normally call Newton's constant 
may be related to the inverse gravitational susceptibility (up to a factor of four) of the background in which all our experiments take place. 

Note that despite the name ``varying-$G$'' in GEVAG, the present framework makes it clear that we should not treat GEVAG as a ``typical'' varying-$G$ theory of the form $G=G(x)$, where $G$ is a field (such as in a scalar-tensor theory of gravity), taking different values at different spacetime points $x$. Rather, our $G_\text{eff}$ depends on the horizon state and on the causal structure of the spacetime. In the field theory of $G(x)$, its value may vary from one spacetime point (or one cosmological epoch) to another, with dynamics that are often governed by an equation of motion. The ``many'' possible values of $G(x)$ are still fundamentally a spacetime variation of a single coupling. The scenario discussed in this work is quite different --- there can be many values of effective gravitational coupling since they are coefficients characterizing how the gravitational system reacts to the variations of different horizons. Notably there is no equation of motion governing the effective gravitational ``constants''.

An important open question is how these distinct horizon-associated responses should enter a consistent bulk gravitational dynamics. One possibility is that they characterize different thermodynamic sectors associated with the respective horizons, without each defining a complete field equation throughout the causal patch. Alternatively, a consistent bulk description may involve a single effective source coupling constructed from the various horizon responses. A more general dynamical completion might instead require additional geometrical or source terms, or relations between the horizon sectors, to accommodate the distinct responses. In the single-horizon GEVAG framework, the effective coupling is determined by the horizon area rather than introduced as an independent scalar field, and the Bianchi identity requires $\nabla^a(G_{\rm eff}T_{ab})=0$ for an Einstein-form field equation with constant $\Lambda$ \cite{2603.23551}. The multi-horizon susceptibility framework, however, does not by itself select among the aforementioned possibilities or determine the corresponding bulk conservation law. Establishing such a dynamical completion therefore remains an interesting open problem.

In multi-horizon spacetimes, the dependence of these susceptibilities on other horizon variables is characterized by the matrix $H_{ij} = \partial \chi_i/\partial A_j$, which quantifies horizon cross-coupling. Our framework thereby avoids the problems of associating a re-scaled gravitational ``constant'' to each horizon when entropy is generalized. In addition, it suggests a possible interpretation of the commonly raised question as to what is the measured Newton constant in laboratory experiments.

The comparison with a dielectric medium is intriguing (though of course not entirely the same). We call the quantity $\chi_i$  gravitational susceptibility based on the analogous susceptibility $\chi_{ij} = \left.\partial P_i/\partial E_j\right|_{\mathbf E=0}$ in the polarization equation
\begin{equation}
P_i = \sum_j \chi_{ij}E_j.
\end{equation}
Then our entropy Hessian matrix $H_{ij}$ is analogous to the differential dielectric response $\partial P_i/\partial E_j$: both describe how a response changes with its arguments. For comparison, a nonlinear dielectric admits higher-order terms\footnote{Another analogy would be a spring that satisfies Hookean law $F=kx$ can nevertheless deform beyond the linear regime and pick up higher order terms: $F=kx+ax^2 + \cdots$.}:
\begin{equation}
P_i = \sum_j \chi_{ij}E_j + \frac{1}{2}\sum_{j,k}\chi_{ijk}E_jE_k + O(E^3),
\end{equation}
where
\begin{equation}
\chi_{ijk} = \left.\frac{\partial^2 P_i}{\partial E_j \partial E_k}\right|_{\mathbf E=0}.
\end{equation}
The susceptibility language therefore makes coupling between horizons something quite natural to contemplate. Indeed, terminology often brings along a conceptual picture and makes some questions seem natural while obscuring others\footnote{{``The limits of my language mean the limits of my world.'' -- Ludwig Wittgenstein, \emph{Tractatus Logico-Philosophicus}.}}.

In Chapter II \emph{Methodology} of Part II. Natural Science in the book ``Philosophy of Mathematics and Natural Science'' by Hermann Weyl \cite{Weyl}, one finds the following paragraph worth quoting:
\begin{quote}
``Since the phenomenological laws are apt to fail wherever the finer internal structure of matter is relevant, the atomic theory must at the same time disclose the limits of their validity and yield the atomic laws which, beyond these limits, take the place of the macroscopic laws. Thus Maxwell had assumed that the electric polarization is proportional to the field strength. This is correct for static and for slowly changing fields, and even for the fields of wireless telegraph which carry out more than a million oscillations per second. But in the domain of the much more rapid optical oscillations we encounter in the new phenomenon of dispersion, the proportionality factor taken as constant by Maxwell --- that is, the constant of dielectricity --- turns out to be dependent on the frequency of oscillation, and this according to laws which are closely connected with the atomic structure of the refracting medium and can only thus be understood.''
\end{quote}
In other words, what we call ``constants'' can turn out to characterize the response of a system under some specified conditions rather than something ontologically fundamental.

Likewise, in our case, the measured Newton constant $G_\text{measured}$ may not be a true constant, though an actual fundamental constant $G$ exists (just like vacuum permittivity exists in electromagnetism). A proposed, but not yet locally derived, interpretation for $G_\text{measured}$ is that it is tied to the cosmological horizon. That is, under this proposed identification,
\begin{equation}
G_\text{measured} = G_c = \left(4\frac{\partial S}{\partial A_c}\right)^{-1}.
\end{equation}
Note that in our symmetric model Eq.(\ref{model}) for example, despite the symmetry under $b \leftrightarrow c$, the fact that $A_c, A_0 \gg A_b$ means that 
\begin{equation}
G_c =\frac{G}{1-\gamma\frac{A_b}{A_0}} \approx G,
\end{equation}
even if
\begin{equation}
G_b =\frac{G}{1-\gamma\frac{A_c}{A_0}}
\end{equation}
can be modified by $A_c$ considerably. 

One virtue for the framework of gravitational susceptibility is that it naturally suggests the possibility of nontrivial coupling between two horizons. When applied to a black hole inside a Universe, the cosmological horizon thus affects the effective gravitational coupling of the black hole, which can be interpreted as an effective mass that is affected by the expansion of the Universe. 
The detailed cosmology requires a careful future study. 

The limiting case $\xi=1$ in the prescribed $\Lambda$CDM example of Sec.\eqref{IV} is nevertheless instructive. It gives $k(a)=3/(1+u)=3\Omega_\Lambda(a)$ and $M_\text{eff}=\mathcal{M}(1+u)/u$, so that a population with conserved comoving number has an \emph{inferred} density $\rho_b\propto a^{-3}+\Omega_{\Lambda0}/\Omega_{m0}$. Thus the model formally exhibits a dust-like contribution together with a constant contribution, while $k\to3$. This is interesting as a scaling correspondence, but it is not a derivation of dark energy: the $\Lambda$CDM history was assumed, the parameter must be tuned to $\xi=1$, and the same limit has $\partial S/\partial A_b\to0$ and $G_b\to\infty$ as $a\to\infty$. The pole is never reached at finite $a$ in this illustrative history, but the effective response becomes unbounded. Future work must test whether a self-consistent cosmological evolution can produce comparable behavior within a regular and physically admissible regime.

More generally, the pole raises an unresolved question already present in single-horizon GEVAG: how should the framework be interpreted when an entropy derivative vanishes? With $S=f(A)/(4G)$, $G_\text{eff}=G/f'(A)$ diverges as $f'(A)\to0$, but this need not entail a curvature singularity. An explicit limiting example follows from the logarithmically corrected FLRW cosmology of Ref.~\cite{2603.23551}. For a negative logarithmic coefficient, write $f'(A)=1-\alpha H^2$ with $\alpha>0$. In the spatially flat, $\Lambda=0$ case with a perfect fluid $p=w\rho$ and constant finite $w$, the Einstein-form equations yield $\rho=3H^2(1-\alpha H^2)/(8\pi G)$ and $\dot H=-\tfrac32(1+w)H^2$ on the branch $f'>0$. As $H^2\to1/\alpha$ from below, $G_\text{eff}\to\infty$ and $\rho\to0$, but $G_\text{eff}\rho\to3/(8\pi\alpha)$, while $H$ and $\dot H$ remain finite. Indeed, $R=3(1-3w)H^2$ and $R_{abcd}R^{abcd}=12H^4[1+(1+3w)^2/4]$ remain finite as well. This regular \emph{limiting} geometry does not establish a physically admissible continuation through the exact zero-susceptibility point, where the inverse-response formula is undefined and the pre-division null-projected relation becomes degenerate. The density-saturation point studied in Ref.~\cite{2603.23551} instead occurs at $f'=1/2$. Whether analogous compensation or regular limiting behavior is possible when one of several horizon susceptibilities vanishes requires a consistent multi-horizon dynamical completion.

In addition, if one includes the dissipative term discussed in Sec.(\ref{V}), i.e., by considering the Debye relaxation analog of gravitational susceptibility, we can expect there to be more interesting and richer effects on the evolution of the Universe. For example, it could potentially give rise to a delayed dark energy, or memory-dependent evolution in the sense that the same instantaneous $A_c$ need not lead to the same $G_b$ if they arrived there through different evolutionary histories. Such history dependence would require internal variables or a nonlocal constitutive law; it does not follow from the instantaneous relation (\ref{ent2}) alone. Numerous proposals involving relaxation in cosmology can already be found in the literature \cite{0606025,1103.1328,1212.4094}. In our case, a delayed cosmological response can potentially arise from non-equilibrium horizon thermodynamics: the gravitational response of one horizon need not adjust instantaneously to changes in another.

Another possible arena to consider the effect of the dissipative term is in the context of the arrow of time \cite{penrose}. The standard argument is that in the very early Universe, gravitational entropy is extremely low. As structures form, gravitational entropy begins to increase. However, Eq.(\ref{ent2}) suggests that gravitational clumping would not \emph{merely} produce gravitational entropy, but rather the entropy production itself would feedback into the strength of the gravitational response, which in turn leads to more clumping. In this way, just like energy becomes a source of gravity, entropy also affects the effective coupling that governs gravitational dynamics. The route is somewhat indirect, however, because Eq.(\ref{ent2}) only deals with internal entropy production of the horizon system, not gravitational entropy production of arbitrary clumped matter. However, since our ``measured $G$'' is associated with the cosmological horizon, ordinary matter clumping inside it can change the geometry of the Universe and thus affects its cosmological horizon (this is of course a sub-leading effect at the perturbation level). The non-equilibrium cosmological horizon then has an associated $\delta_\text{int}S_c$, thus allowing horizonless structures to affect the ``measured $G$'' indirectly. A fully specified irreversible dynamics would be needed before this possibility could be assessed quantitatively. Whether this will ameliorate the arrow of time problem or worsen it remains to be seen, though I do not expect it to completely resolve the problem without imposing the Past Hypothesis \cite{Albert}.

\begin{acknowledgments}
This research is supported by NUAA funding No.1018-ILF26028.
\end{acknowledgments}

\end{document}